\documentclass[twocolumn, aps]{revtex4}
\usepackage[english]{babel}
\usepackage{amsmath}
\usepackage{amssymb}
\usepackage[dvips]{graphicx}
\usepackage{psfrag}

\newcommand{\bs}[1]{\boldsymbol{#1}}

\begin{document}

\title{Steady state particle distribution of a dilute sedimenting suspension}

\author{Bogdan Cichocki}

\author{Krzysztof Sadlej}
 \email{Krzysztof.Sadlej@fuw.edu.pl}
\affiliation{
Warsaw University, Institute of Theoretical Physics\\
Ho\.za 69, 00-681 Warsaw, Poland
}

\date{\today}
\begin{abstract}
Sedimentation of a non-Brownian suspension of hard particles is studied. It is shown that in the
low concentration limit a two-particle distribution function ensuring finite particle correlation length 
can be found and explicitly calculated. The sedimentation coefficient is computed.
Results are compared with experiment.
\end{abstract}
\maketitle

\section{Introduction}
\label{sec:sec001}
In recent years the familiar problem of sedimentation of a non-Brownian suspension of particles 
has gained interest as new insight into the phenomenon is gathered. Despite vivid 
research in the field, some questions still remain intriguing puzzles.  
Even when considering observation time scales and dimension regimes, which sanction the
neglect of Brownian motion, the dynamics show to be many-body and highly
chaotic leading to astonishing consequences.

One of the main, still open problems, is the determination of the particle probability distribution function
for a steadily sedimenting suspension. It was first challenged by  Batchelor \cite{viii}. In order to derive
an equation for the pair-distribution function he considered a polidisperse suspension. Taking then the limit
of identical particles, a particle distribution function for mono-disperse suspensions could be derived. But these
results where shown to be ambiguous (depending on the way the limit is taken) and therefore doubtful (\cite{viii}, 
\cite{ix}, \cite{iv}).
The problem of particle distribution is often skimmed over
by the assumption, that the particles are randomly distributed in the fluid. Indeed this will be the 
case, if Brownian motions play a significant role in the dynamics of the system, creating a randomising mechanism. 
But whenever this type of stochastic motion is absent, there seems to be no justification for such assumptions.
Further, it has been \cite{vii} pointed out, that a equilibrium (equal probability) distribution 
of particles results in characteristics of a non-Brownian suspension, 
which have not been detected in experiments \cite{iii}. 
Moreover it can actually be explicitly shown that such a distribution cannot be a solution 
of the Liouville equation incorporating full multi-body interactions, therefore falling completely
out of consideration.  

In this work we propose a way to tackle this issue. The derivation of the statistical properties of the system 
will be based on the conjecture that correlations in steady state must be of finite
length. This assumption, originating in the basic ideas of statistical mechanics, has profound consequences.
It, as will be shown, leads to an effective equation for the pair distribution function.  
Truncation of long range hydrodynamic interactions will be achieved through a screening mechanism similar in spirit
to that introduced by Koch\&Shaqfeh \cite{v}. 

\section{BBGKY hierarchy}
The system under consideration is a suspension of $N$ spheres of equal radius $a$ immersed in 
an incompressible fluid of shear viscosity $\eta$. In the regime studied the particle Reynolds
number is assumed to be small enough to treat inertial effects as virtually absent. Moreover the
Peclet number describing the relative influence of hydrodynamic effects to Brownian motion,
is presumed to be large. On those assumptions and on the laboratory time scales imposed 
the fluid motion is governed by the stationary Stokes equation. The dynamics of the particles are therefore
given by the equations of motion
	\begin{equation}
	\label{eq:eq000}
		\frac{d\bs{R}_i}{dt}=\bs{U}_i = \sum_{j=1}^{N}\bs{\mu}_{ij}(\bs{X})\bs{F}_j
	\end{equation}
where $\bs{\mu}$ is the translational part of the 
mobility matrix, principally a matrix function of the configuration 
$\bs{X}=(\bs{R}_1,\ldots,\bs{R}_N)$ of all the particles (if not indicated otherwise, dimensionless - 
normalised to $2a$ - distances will be used). It describes the hydrodynamic 
interactions between particles and connects the forces acting on particles to the velocities
they acquire in a given configuration.
The forces $\bs{F}_j$ will be assumed to be constant and equal $\bs{F}$ for all particles.

The particle distribution function satisfies the Liouville equation, 
which for the system considered, has the following structure
	\begin{equation}
		\frac{\partial \rho(\bs{X};t)}{\partial t}+\sum_{i,j}\nabla_i \cdot
                     \left[ \bs{\mu}_{ij}(\bs{X})\bs{F}\rho(\bs{X};t)\right]=0
	\end{equation}
The mobility matrix can be found following a formal procedure presented for example in ref.\cite{i}. Summarising, 	
in order to calculate the hydrodynamic interaction, the boundary conditions on the particles surfaces are substituted by
induced force densities introduced for each particle.
These force densities, induced by fluid flow, produce subsequent fluid velocity fields which are propagated and 
reflect off other particles and thereby cause interactions. 
In consequence, the  mobility matrix can be expressed as a scattering sequence equivalent to a superposition of 
all these reflections. This sequence contains one-particle scattering operators, which construct induced 
force densities on particles given velocity fields around them, and Green 
operators which propagate the influence of a force density concentrated on a given particle, resulting in 
fluid velocity fields around other particles. In an unbounded, infinite system, the Green operator is the Oseen tensor.

The full $N$-body mobility matrix can therefore be expanded in terms of the number of particles which enter each 
term of the scattering sequence. This so called cluster expansion has the structure
\begin{eqnarray}
\bs{\mu}_{11}(X)=\bs{\mu}_{11}^{(1)}(1)+\sum_{i=2}^{N} \bs{\mu}_{11}^{(2)}(1i)+\ldots
                                      \nonumber\\
\bs{\mu}_{12}(X)=\bs{\mu}_{12}^{(2)}(12)+\sum_{i=3}^{N} \bs{\mu}_{12}^{(3)}(12i)+\ldots
\end{eqnarray}
where $\bs{\mu}^{(s)}_{ij}(1\ldots,s)$ denotes all the terms of the scattering sequence of $\bs{\mu}$
which involve only, but all the particles $\{1,\ldots,s\}$.

Next we introduce reduced distribution functions
 \begin{equation}
   n(1,2,\ldots,s)  =  \frac{N!}{(N-s)!}\int d(s+1)\cdots dN \rho(X)
 \end{equation}
These functions represent the average densities of pairs, triplets, etc. of particles in given configurations.

An integration of the Liouville equation in all the variables except the first $s$, together with the cluster expansion
of the hydrodynamic interactions, leads to an infinite hierarchy of equations governing the time evolution of the
reduced distribution functions. This is the analogue of the BBGKY (Bogolubov-Born-Green-Kirkwood-Yvon) hierarchy introduced
in the kinetic theory of gases \cite{x}. It should be
considered in the thermodynamic limit when the number of particles and the volume of the system go to infinity, while the
density is kept constant. The equations for the two and three particle distribution functions have the structure
{\setlength\arraycolsep{2pt}  
\begin{eqnarray}
\label{eq:eq001}
\frac{\partial n(12)}{\partial t} & = & 
               -\sum_{i,j=1,2}\nabla_i\cdot \bs{\mu}_{ij}(12)\bs{F}n(12)
               \nonumber\\
               &&-\int d3\sum_{i=1,2}\nabla_i\cdot[\bs{\mu F}]_i(12;3)n(123)+\ldots
\end{eqnarray}}
where for example
\begin{equation}
	 [\mu \bs{F}]_1(12;3)=[\mu_{11}^{(2)}(13)+\mu_{13}^{(2)}(13)+
                          \sum_{j=1}^{3}\mu_{1j}^{(3)}(123)]\bs{F} 
\end{equation}
and
\begin{equation}
\label{eq:eq005}
\frac{\partial n(123)}{\partial t}=-\sum_{i,j=1}^{3}\nabla_i\cdot\bs{\mu}_{ij}(123)\bs{F}n(123) +\ldots
\end{equation}
where in both equations, terms depending on higher-order distribution functions have been omitted.

The reduced distribution functions factorize for groups of particles that are far away from each other.
Therefore these functions can be expanded in terms of correlation functions according to   
	\begin{eqnarray}
     \label{eq:eq002}
		n(12) & = & h(1)h(2)+h(12)
		\nonumber\\
		n(123) & = & h(1)h(2)h(3)+ h(12)h(3)
                 \nonumber\\
                  &&+h(13)h(2)+h(23)h(1)+h(123)
	\end{eqnarray} 
where $h(1\ldots s)$ is a s-particle correlation function which vanishes whenever any subset 
of particles is dragged away from the rest. If the correlations in a system are to be of 
finite length, which will be assumed, the correlation functions must 
decay faster then the  inverse of the inter-particle distance cubed.

Using the expansion (\ref{eq:eq002}), the BBGKY hierarchy for the reduced distribution functions can be 
transcribed into a hierarchy of equations for the time evolution of the successive correlation functions. 
An analysis of these equations shows that if system configuration is such
that particles form two uncorrelated clusters, some expressions on the r.h.s, formulated in terms of the scattering sequence,
contain single Green operators (so called connection or articulation lines \cite{i},\cite{ii}) 
joining these two groups of particles. Such terms result
in long range contributions because a solitary Green operator includes slowly decaying
parts proportional to $1/r^{\gamma}$ where $r$ is the relative distance between the groups and $\gamma=1,2,3$. The evolution
of a correlation function is therefore given by an equation which contains long range terms and is therefore non-local 
in space. This structure of the hierarchy equations is inconsistent with the finite correlation length assumed
for the correlation functions. As will be shown in the next section, considering the low concentration limit
a particle distribution can be found, which cancels out all long range terms in the hierarchy, thereby saving finite
correlation length.
 
\section{Low concentration limit pair distribution function}
In the low concentration limit the dominating part of a $s$-particle reduced distribution function is 
proportional to $n^s$. Consequently, the first terms of the hierarchy equations 
have a well established order in density if the concentration of particles is low. 

Consider the equation giving the time evolution of
the two-particle correlation function. It is derived through a substitution of (\ref{eq:eq002})
into (\ref{eq:eq001}). In the lowest order of density it vanishes due to the relation ~\cite{viii}
\begin{equation}
	\sum_{i,j=1,2}\nabla_i\cdot\bs{\mu}_{ij}(12)=0.
\end{equation}
which is a consequence of the isotropic nature of the hydrodynamic interactions.
The two-particle equation yields therefore in the dilute limit no condition for the two-particle correlation function. 
This peculiarity was also encountered by Bathelor~\cite{viii}. His solution was based on the analysis of a polidisperse 
suspension where the two-particle equation yields a condition, but was shown to be ambiguous (\cite{viii}
,\cite{ix},\cite{iv}).  
We turn our attention to the next hierarchy equation, i.e. the third equation which emerges when (\ref{eq:eq002}) 
is addressed to (\ref{eq:eq005}), truncated to the terms proportional to the density cubed.
   
As pointed out, the r.h.s of this equation contains long range terms.
If  all three particles are far away from each other, both sides of the
equation contain only shot range terms. But consider the case when one particle (e.g. the particle with index~$1$) is far away 
from a close pair of particles (consequently particles with indices~$2$ and~$3$). All the terms,
which in such a configuration lead to long range contributions can be identified and singled out 
resulting in the expression
{\setlength\arraycolsep{2pt}  
\begin{eqnarray}
\label{eq:eq003}
    n\bs{\nabla}_2\cdot\big[(\bs{\mu}_{21}(2|1)&+&\bs{\mu}_{21}(23|1)+\bs{\mu}_{21}(2|3|1))n(23)\big]+
    \nonumber\\
    n\bs{\nabla}_3\cdot\big[(\bs{\mu}_{31}(3|1)&+&\bs{\mu}_{31}(32|1)+\bs{\mu}_{31}(3|2|1))n(23)\big]
\end{eqnarray}}
where each vertical line stands for a single Green operator connecting uncorrelated
particles. For example $\bs{\mu}_{21}(23|1)$ stands for the sum of all terms of the scattering sequence of
$\bs{\mu}^{(3)}_{21}(123)$ that contain only a single Green operator connecting particle 1 and the group~(2,3). 

In what follows, we will show that there exists a two-particle probability distribution function which cancels out
these long range terms and therefore leads to finite correlation length

In equation (\ref{eq:eq003}) the long range contributions  appear in the Green operators binding
particle $1$ with the group consisting of particles $2$ and $3$.
Their explicit form is retrieved through an analysis of the multipole matrix elements of the Green operators,
which can be found e.~g. in~\cite{ii}.

When taking into account the symmetry relations upon exchange of particles,
and a change of variables where $\bs{r}$ becomes the vector joining the distant particle 
and the center of the group of close particles, while $\bs{R}$ the vector denoting the relative 
position of the two particles within the group,  
we reach the conclusion that most of the long range contributions cancel out automatically - only the multipoles 
proportional to the inverse of the inter-particle distance survive.

A further expansion in terms of $R=|\bs{R}|$ ($R\ll r$) shows that only terms
proportional to $1/r^2$ remain. The equation which emerges has an identical
structure to the one derived by Batchelor\&Green~\cite{xi} 
and describing the relative motion of a pair of particles in a shear flow. 
The resulting particle distribution function $n(\bs{R})=n^2g(R)$ 
is isotropic. It is given by the solution of the 
differential equation
\begin{equation}
        \left\{\left(1-A\right)\frac{1}{g}\frac{dg}{dR} -  \frac{3(A-B)}{R}
            -\frac{dA}{dR} \right\} = 0
\end{equation}
$A$ and $B$ are functions of the scalar distance $R$ given by the relations
$A=2(\alpha_{11}^{td}+\alpha_{12}^{td})/R$ and $B=2(\beta_{11}^{td}+\beta_{12}^{td})/R$ where
$\alpha_{ij}^{td}$ and $\beta_{ij}^{td}$ are functions appearing in the explicit form
of the generalised two-particle mobility matrix connecting the translational velocity of the particles to the 
stresslet~\cite{vi}.
Since $\lim_{R\to\infty}g(R) = 1$ 
\begin{equation}
  \label{eq:eq004}
 g(R)=\frac{1}{1-A}\exp\left[\int_{R}^{\infty} \frac{3(B-A)}{R(1-A)}dR\right]
\end{equation}
Note that this solution is independent of the direction distinguished by 
the external force field, the effects of which are in the low concentration limit completely dispersed by the isotropic
hydrodynamic interactions. Further, to verify the consistency of the reasoning imposed, 
it can be shown that the two-particle distribution function derived
(\ref{eq:eq004}) guarantees convergence of all terms up to order $n^3$ in equation (\ref{eq:eq001}) governing 
the time evolution of the two-particle reduced distribution function.

\section{Results}
\subsection{Stationary particle distribution function}
\begin{figure}
  \psfrag{gRR}{\small{$g(R)$}}
  \psfrag{R}{\small{$R$}}
   \begin{center}
     \caption{The two-particle reduced distribution function $g(R)$ satisfying the finite correlation length
              criterion}
      \label{fig1}
     \includegraphics[scale=0.5]{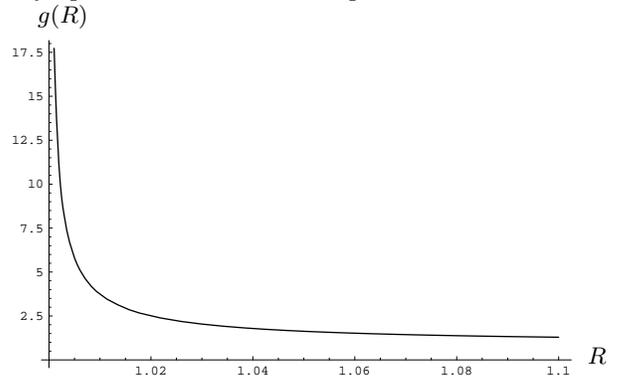}
   \end{center}
   \end{figure}

  The functions $A(R)$ and $B(R)$ can be expressed as series in inverse powers of inter-particle distance~\cite{vi}.
 In numerical approximations of the integral (\ref{eq:eq004}) these sequences are truncated at 1000 and
 800 terms respectively. Lubrication corrections and far limit asymptotic with $g(R)-1\approx 0.1953/R^6$
 are taken into account. The computed function $g(R)$ is plotted at fig.~\ref{fig1}. Further, the structure factor at
 $\bs{k}=\bs{0}$ is calculated: $S(\bs{0})=1-1.64\phi$. 

\subsection{Sedimentation coefficient}
\begin{figure}
   \begin{center}
     \caption{Dimensionless sedimentation velocity as a function of volume fraction - experimental results
       after Hanratty et al.\cite{xii}. Points represent experimental data and the solid curve is a fit.
       The solid line represents the sedimentation coefficient $K=-3.87$, whereas the dotted line is the coefficient
       for a suspension in equilibrium with $K=-6.55$. }
      \label{fig2}
     \includegraphics[height=9cm, width=8.5cm]{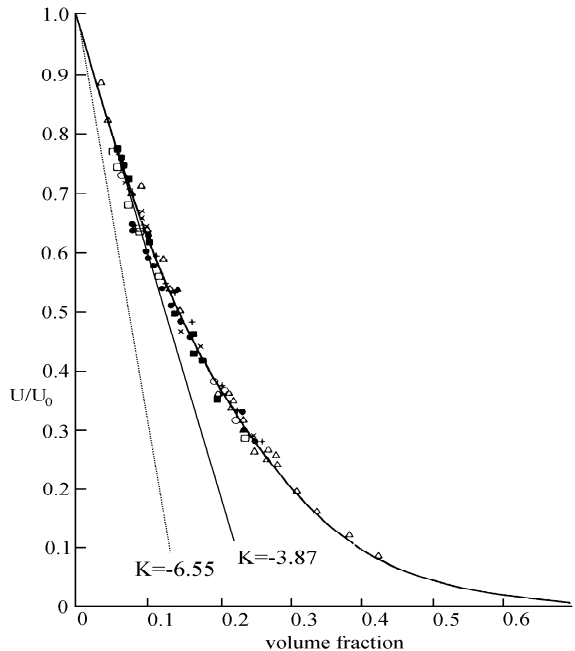}
   \end{center}
   \end{figure}

In the low concentration limit the average sedimentation velocity $U$ of a particle in the suspension measured relative
to the Stokes velocity $U_0$ may be expanded in a series in the powers of volume fraction $\phi =4\pi a^3n/3$.
 \begin{equation}
   \frac{U}{U_0}=1+\phi K+\mathcal{O}(\phi^2)
 \end{equation}
The linear coefficient $K$ can be expressed in terms of a microscopic 
expression ~\cite{xiii} involving two-body mobility matrices and the pair distribution function.
\begin{equation}
	K  =  K_{0}+
          \frac{2}{\pi\mu_0}\int_{R_{12}\geq 1}(g(12)-1)\textrm{Tr}\left[\sum_{i=1}^{2}\bs{\mu}_{1i}^{(2)}(12)\right]
          d\bs{R}_{12}
    \end{equation}
where $K_{0}=-6.546$~(\cite{xiii}, \cite{xiv}) is the sedimentation coefficient for a equilibrium distribution of hard spheres 
and $\mu_0$ is the mobility of a single sphere. Dimensionless distance $R_{12}$ normalised to $2a$ is adopted.
The integral can be interpreted as a correction emerging due to the change of the distribution
away from equilibrium.
A numerical calculation yields the result $K=-3.87$. It is compared with experimental 
data of Hanratty et al. on fig.~\ref{fig2}. A fit to 
the experimental data of Ham et al.~\cite{xv} leads to the sedimentation coefficient equal to $-3.9$,
in excellent agreement with the calculated value.

\section{Concluding remarks}
	We have shown that a mono-disperse suspension of hard, non-Brownian particles 
can develop a micro-structure which insures finite correlation length. A scheme 
based on the formal schemes of non-equilibrium statistical mechanics and a detailed
analysis of hydrodynamic interaction was used in the low-concentration regime
to derive an equation for the two-particle
distribution function. The solution shows to be isotropic due to the domination of the isotropic
hydrodynamic interactions. The mechanism which insures the cutoff of long range correlations 
can be interpreted as hydrodynamic screening in the sense that the micro-structure of the suspension
arranges itself in a way which truncates long range parts of the interactions. A close pair of 
particles feels the influence of a third distant particle through an effective shearing flow. Screening 
leads to a configuration of the close pair, which hinders the effect of the distant particle.
The idea, that a suspension might develop a microstucture which changes the characteristics of it, 
was introduced by Koch\&Shaqfeh \cite{v} 
upon the analysis of the problem of diverging velocity fluctuation.    

Further, the sedimentation coefficient was calculated. The value $-3.87$ was found. It
agrees very well with the experimental results cited.

\end{document}